\documentclass[12pt]{article}
\usepackage[small]{titlesec}
\usepackage{hyperref}
\usepackage{amssymb,amsmath,amsthm}
\usepackage{enumerate}
\usepackage{cite}
\usepackage[letterpaper,hmargin=2.70cm,vmargin=3.2cm]{geometry}

\usepackage{setspace}
\newtheorem{theorem}{Theorem}[section]
\newtheorem{proposition}[theorem]{Proposition}
\newtheorem{lemma}[theorem]{Lemma}
\newtheorem{corollary}[theorem]{Corollary}
\theoremstyle{definition}
\newtheorem{definition}[theorem]{Definition}

\theoremstyle{remark}
\newtheorem*{remark}{Remark}
\newtheorem*{acknowledgments}{Acknowledgments}

\newcommand{\abs}[1]{\left|#1\right|}
\newcommand{\norm}[1]{\left\|#1\right\|}

\newcommand{\inner}[2]{\left\langle#1,#2\right\rangle}

\newcommand{\cH}{{\cal H}}
\newcommand{\cB}{{\cal B}}

\newcommand{\cc}[1]{\overline{#1}}
\renewcommand\tilde{\widetilde}
\newcommand{\ourclass}{\text{Sym}^{(1,1)}_\text{R}({\cal H})}

\numberwithin{equation}{section}

\DeclareMathOperator{\im}{Im}
\DeclareMathOperator{\dom}{Dom}
\DeclareMathOperator{\Ker}{Ker}

\DeclareMathOperator{\ran}{Ran}
\DeclareMathOperator{\Sp}{Spec}
\DeclareMathOperator{\Span}{Span}

\DeclareMathOperator{\assoc}{Assoc}

\begin{document}
\begin{titlepage}
\title{On the spectral characterization of entire operators with
	deficiency indices $(1,1)$%
\thanks{%
Mathematics Subject Classification(2000):
46E22, 
47A25, 
47B25, 
47N99.} 
\thanks{%
Keywords: symmetric operators, entire operators, de Branges spaces, spectral
	analysis.}
\\[6mm]}
\author{
\textbf{Luis O. Silva}
\\
\small Departamento de M\'{e}todos Matem\'{a}ticos y Num\'{e}ricos\\[-1.6mm]
\small Instituto de Investigaciones en Matem\'aticas Aplicadas y
	en Sistemas\\[-1.6mm]
\small Universidad Nacional Aut\'onoma de M\'exico\\[-1.6mm]
\small C.P. 04510, M\'exico D.F.\\[-1.6mm]
\small \texttt{silva@leibniz.iimas.unam.mx}
\\[4mm]
\textbf{Julio H. Toloza}\thanks{Partially supported by CONICET
	(Argentina) through grant PIP 112-200801-01741}
\\
\small IMIT--CONICET\\[-1.6mm]
\small Universidad Nacional del Nordeste\\[-1.6mm]
\small Avenida Libertad 5400, W3404AAS Corrientes, Argentina\\[-1.6mm]
\small \texttt{{jtoloza@exa.unne.edu.ar}}}
\date{}
\maketitle
\begin{center}
\begin{minipage}{5in}
\centerline{{\bf Abstract}} \bigskip
For entire operators and entire operators in the generalized sense, we
provide characterizations based on the spectra of their selfadjoint
extensions. In order to obtain these spectral characterizations, we
discuss the representation of a simple, regular, closed symmetric
operator with deficiency indices $(1,1)$ as a multiplication operator
in a certain de Branges space.
\end{minipage}
\end{center}
\thispagestyle{empty}
\end{titlepage}

\section{Introduction}

M. G. Krein introduced the concept of entire operators in the search
of a unified treatment of several classical problems in analysis
\cite{krein1,krein2,krein3,krein4}; a review book on this matter is
\cite{gorbachuk}. In general, it may be difficult to determine whether
a given operator is entire due to the lack of criteria not based on
finding the entire gauge. In this work we present necessary and
sufficient conditions for an operator to be entire for the case of
deficiency indices $(1,1)$. These conditions are based exclusively on
the distribution of the spectra of two von Neumann selfadjoint
extensions of the operator. More concretely (the precise statement is
Theorem~\ref{cor:spectrum-tells-if-operator-is-entire}):\\[4mm]
{\em Let $A$ be a simple, regular, closed symmetric operator with deficiency
  indices $(1,1)$. Consider two of its selfadjoint extensions $A_0$ and
  $A_{\gamma}$. Then $A$ is entire if and only if
  $\Sp(A_0)$ and $\Sp(A_{\gamma})=\{x_n\}$ obey the following
  conditions:
\begin{itemize}
\item The limit
	$\displaystyle{\lim_{r\to\infty}\sum_{0<|x_n|\le r}
		\frac{1}{x_n}}$
	exists;
\item $\displaystyle{\lim_{n\to\infty}\frac{n}{x_n^{+}}
		=\lim_{n\to\infty}\frac{n}{x_n^{-}}<\infty}$;

\item Assuming that $\Sp(A_{\beta})=\{b_n\}$, define
	\[
	h_\beta(z):=\left\{\begin{array}{ll}
			\displaystyle{\lim_{r\to\infty}\prod_{|b_n|\le r}
			\left(1-\frac{z}{b_n}\right)}
				& \text{ if }0 \not\in\Sp(A_{\beta}),
			\\
			\displaystyle{z\lim_{r\to\infty}\prod_{0<|b_n|\le r}
			\left(1-\frac{z}{b_n}\right)}
				& \text{ otherwise. }
			   \end{array}\right.
	\]
	The series
	$\displaystyle{
		\sum_{n\in\mathbb{N}}\abs{\frac{1}
		{h_{0}(x_n)h_{\gamma}'(x_n)}}}$ is convergent.
\end{itemize}
Here $\{x_n^+\}$ and $\{x_n^-\}$ are, respectively, the sequences of
positive and negative elements of $\Sp(A_{\gamma})$.
}\\[4mm]
\indent Our spectral characterization was motivated by a result, due
to Woracek \cite{woracek}, which gives necessary and sufficient
conditions for a de Branges space to have 1 as an associated function
(Theorem 3.1). These conditions are formulated in terms of the spectra
of two particular selfadjoint extensions of the multiplication
operator in the de Branges space. On the basis of a simple result
(Lemma 3.4) we reformulate the necessary and sufficient conditions in
terms of two arbitrary selfadjoint extensions
(Proposition~\ref{thm:1-is-assoc-to-B-extended}). These results are
then combine with spectral theory and de Branges spaces theory to
obtain necessary and sufficient conditions for 1 to be in the de
Branges space (Proposition~\ref{prop:1-in-dB-boosted}).

A recent general result \cite{woracek2} gives, as a particular case,
necessary and sufficient conditions for a de Branges space to contain
the function 1. The present work presents an alternative approach to this
question and, by means of Lemma~\ref{lemma:second}, provides a way 
for reformulating the necessary and sufficient conditions of \cite{woracek2} 
in more general terms.

Having obtained the results mentioned above on de Branges spaces, we
establish the spectral characterization of entire operators
(Theorem~\ref{cor:spectrum-tells-if-operator-is-entire}) on the basis
of the representation of any regular, simple, symmetric operator, with
deficiency indices $(1,1)$, as the multiplication operator in a
certain de Branges space (Section~4). The realization of this representation
parallels the construction by Krein, even though the latter yields a de
Branges space only when the operator is entire. We ought to mention that
an alternative representation theory was developed recently
in \cite{martin}, although based on a quite different approach. 

Besides the necessary and sufficient conditions for the spectra of two
selfadjoint extensions of an entire operator, we also provide the
spectral characterization for operators that are entire in the generalized 
sense (Theorem~\ref{thm: a-is-entire-generalized}). The key ingredient of 
this characterization (Proposition~\ref{prop:assoc-of-hat-h}) is 
the representation developed in Section~4.

In the process of deriving the results of this work, we touch upon the
treatment of some two-spectra inverse problems involving selfadjoint
extensions of the symmetric operators considered here
(Corollary~\ref{prop:inverse-problem}). This matter will be consider in
a forthcoming paper.  It is worth remarking that the theory of de
Branges spaces has been already applied to inverse spectral problems
for Schr\"odinger operators \cite{remling1,remling2}.

A short review on entire operators with deficiency indices $(1,1)$ is
given in Section~2. As a by-product, we provide a proof of a result by
Krein related to the properties of gauges
(Proposition~\ref{proposition:good-gauge}). Section~3 starts with a
brief account of some facts on de Branges spaces.

\begin{acknowledgments}
  The authors sincerely thank the anonymous referee for drawing their
  attention to \cite{martin} and \cite{woracek2}, and also for his
  comments and suggestions which led to an improved presentation of
  this work and to the inclusion of new material.  J.\ H.\ T.\ thanks
  IIMAS--UNAM for their hospitality, where part of this work was
  done. He also thanks CONICET for providing partial financial
  support.
\end{acknowledgments}

\section{Review on entire operators}
Let $\text{Sym}^{(1,1)}_\text{R}({\cal H})$ denote the class of
simple, regular, closed symmetric operators densely defined on a
Hilbert space $\cH$, whose deficiency indices are $(1,1)$.  It is
known, and easily verifiable, that because of the simplicity this
class is not empty only when $\cH$ is separable \cite[Section
2]{krein4}.  The selfadjoint extensions of a given operator
$A\in\text{Sym}^{(1,1)}_\text{R}({\cal H})$ shall be denoted by
$A_\beta$ with $\beta\in[0,\pi)$ (other parametrizations may be chosen
as well). The spectra of such selfadjoint extensions are always
discrete and of multiplicity one. Also, the spectra of any two
selfadjoint extensions interlace and, moreover, every point of the
real line belongs to the spectrum of a unique selfadjoint extension.

$\text{Sym}^{(1,1)}_\text{R}({\cal H})$ as a set is invariant under
similarity transformations: If $V:\cH\to\cH'$ is one-one and onto,
then $\text{Sym}^{(1,1)}_\text{R}({\cal H'})=
V\text{Sym}^{(1,1)}_\text{R}({\cal H})V^{-1}$ (where the domains of
operators are transformed accordingly).

In what follows, the inner product will be assumed anti-linear in the first 
argument.

Given $A\in\text{Sym}^{(1,1)}_\text{R}({\cal H})$, a vector
$\mu\in\cH$ is called a {\em gauge} for $A$ if
\begin{equation}\label{eq:gauge-definition}
\cH = \ran(A-z_0I)\dot{+}\Span\{\mu\},
\end{equation}
for some $z_0\in\mathbb{C}$. Once a gauge has been chosen, we look for
the set of complex numbers for which (\ref{eq:gauge-definition}) fails
to hold:
\[
S_\mu = \mathbb{C}\setminus\left\{z\in\mathbb{C}:
	\cH = \ran(A-zI)\dot{+}\Span\{\mu\}\right\}.
\]
That is, $w\in S_\mu$ if and only if $\mu\perp\Ker(A^*-\cc{w}I)$.
Likewise, for every selfadjoint extension $A_\beta\supset A$ we define
\[
S_\mu^{(\beta)} = \left\{z\in\mathbb{C}\setminus\Sp(A_\beta):
	\inner{\mu}{\psi^{(\beta)}(\cc{z})}=0 \right\},
\]
where
\begin{equation}
  \label{eq:cayley}
  \psi^{(\beta)}(z)
	:=\left(A_\beta-z_0I\right)\left(A_\beta-zI\right)^{-1}\psi_0
\end{equation}
is the generalized Cayley transform of
$\psi_0\in\Ker(A^*-z_0I)$. Since
$\psi^{(\beta)}(z)\in\Ker(A^*-zI)$ for every
$z\in\mathbb{C}\setminus\Sp(A_\beta)$ (see \cite{gorbachuk}), we
conclude that
\[
S_\mu^{(\beta)}\subset S_\mu \subset S_\mu^{(\beta)}\cup\Sp(A_\beta).
\]
It follows that $S_\mu$ is at most a countable set with no finite accumulation
points. Also, it is easy to verify that
\[
S_\mu = \bigcup_{\beta} S_\mu^{(\beta)}.
\]
For symmetric operators of the class considered here, one can easily
give a gauge such that the exceptional set $S_\mu$ lies entirely on
the real line.

\begin{lemma}\label{lemma:non-orthogonality}
Let $A\in\text{Sym}^{(1,1)}_\text{R}({\cal H})$. Take an
eigenvector $\mu_0$ of some selfadjoint extension $A_\beta$ as a gauge
for $A$.
Then $S_{\mu_0}=\Sp(A_\beta)\setminus\{x_0\}$, where $x_0$ is
the eigenvalue associated to $\mu_0$.
\end{lemma}
\begin{proof}
The inclusion $\Sp(A_\beta)\setminus\{x_0\}\subset S_{\mu_0}$ is
straightforward, thus we shall only deal with the converse
inclusion.  Since necessarily $x_0\not\in S_{\mu_0}$, it suffices to
show that $S_{\mu_0}$ is contained in $\Sp(A_\beta)$. Suppose that
$v\in\mathbb{C}\setminus\Sp(A_\beta)$. For such a $v$,
$\psi^{(\beta)}(v)=\left(A_\beta-z_0I\right)\left(A_\beta-vI\right)^{-1}\psi_0$
is well defined, where we have chosen $z_0$ non real such that
$\psi_0\in\Ker(A^*-z_0I)$ is not orthogonal to $\mu_0$. Then,
\begin{equation*}
\inner{\mu_0}{\psi^{(\beta)}(v)}
	= \inner{\left(A_\beta-\cc{v}I\right)^{-1}
		\left(A_\beta-\cc{z_0}I\right)\mu_0}{\psi_0}
	= \frac{x_0-z_0}{x_0-v}\inner{\mu_0}{\psi_0}
	\neq 0,
\end{equation*}
thus implying $v\not\in S_{\mu_0}$.
\end{proof}

This result allows us to prove the following
assertion first formulated by Krein without proof in \cite[Theorem 8]{krein2}.

\begin{theorem}[Krein]\label{proposition:good-gauge}
For every $A\in\text{Sym}^{(1,1)}_\text{R}({\cal H})$, there exists a
gauge $\mu$ such that $S_\mu\cap\mathbb{R}=\emptyset$.
\end{theorem}
\begin{proof}
  Choose $\mu_1$ and $\mu_2$ among the eigenstates of two different
  selfadjoint extensions, say, $A_\beta\mu_1=x_1\mu_1$ and
  $A_{\beta'}\mu_2=x_2\mu_2$ ($x_1\ne x_2$). By
  Lemma~\ref{lemma:non-orthogonality} we have
  $S_{\mu_1}=\Sp(A_\beta)\setminus\{x_1\}$ and
  $S_{\mu_2}=\Sp(A_{\beta'})\setminus\{x_2\}$, therefore
\begin{equation}\label{eq:empty-set}
S_{\mu_1}\cap S_{\mu_2}=\emptyset,
\end{equation}
due to the disjointness property mentioned above.

Next, we show that for every
$x\in\mathbb{R}\setminus S_{\mu_2}$ there
exists an unique $g(x)\in\mathbb{C}$ such that
$\left(\mu_1+g(x)\mu_2\right)\perp\Ker(A^*-xI)$: The uniqueness
follows from the fact that $x\not\in S_{\mu_2}$.  As for the existence,
choose some selfadjoint extension $A_{\beta''}$ such that
$x\not\in\Sp(A_{\beta''})$ and set
\[
g(x) = -\frac{\inner{\psi^{(\beta'')}(x)}{\mu_1}}
		{\inner{\psi^{(\beta'')}(x)}{\mu_2}}.
\]

Finally, choose some $\tilde{g}\in G:=\mathbb{C}\setminus\{g=g(x):
x\in\mathbb{R}\setminus S_{\mu_2}\}$.  The set $G$ is non empty since
$g(x)$ is differentiable in any closed interval of $\mathbb{R}\setminus
S_{\mu_2}$ and it cannot produce a space filling curve \cite[Section
5.4]{sagan}.  Then $\mu=\mu_1+\tilde{g}\mu_2$ satisfies the claimed
property as a result of (\ref{eq:empty-set}).
\end{proof}

\begin{remark}
Lemma~\ref{lemma:non-orthogonality} implies that eigenstates of
two different selfadjoint extensions, of an operator in
$\text{Sym}^{(1,1)}_\text{R}({\cal H})$, are never orthogonal to one another.
This is of course expected.
\end{remark}

By Theorem~\ref{proposition:good-gauge}, for operators of the class
considered here it is always possible to find a gauge (indeed
an uncountable number of them) such that the exceptional set $S_\mu$ lies
outside the real line. A distinguished class of operators is the following one.

\begin{definition}\label{def:entire-operator}
An operator $A\in\text{Sym}^{(1,1)}_\text{R}({\cal H})$
is called {\em entire} if there exists a gauge $\mu\in\cH$ such that
$S_\mu=\emptyset$, in which case $\mu$ is said to be an {\em entire gauge}.
\end{definition}

In order to decide whether an operator is entire, the notion of
universal directing functional is sometimes useful. The following
statement is classical \cite[Chapter 2]{gorbachuk}.

\begin{proposition}
Suppose $A\in\text{Sym}^{(1,1)}_\text{R}({\cal H})$ admits a universal
directing functional $\Phi(\cdot,z)$ such that, for some $\mu\in\cH$,
\begin{equation}\label{eq:condition-universal-gauge}
\Phi(\mu,z)\neq 0 \quad\text{for all}\quad z\in\mathbb{C}.
\end{equation}
Then $A$ is entire and $\mu$ is an entire gauge for $A$.
Conversely, associated to an entire operator
with entire gauge $\mu$, there is an universal directing functional
that satisfies (\ref{eq:condition-universal-gauge}).
\end{proposition}

We note that the determination of a universal directing functional may
not be easier than to find an entire gauge by brute force.  One of the
aims of the present work is to provide alternative criteria for
determining when an operator in $\ourclass$ is entire. As we already
have mentioned, these criteria will rely upon the distribution of
elements of the spectra of selfadjoint extensions, thus not requiring
the searching of gauges with particular properties. However, we
should mention that the spectral characterization discussed in the
present work may not necessarily be easier to use in practice.

\section{Some known and new results on de Branges spaces}

Let ${\cal B}$ denote a nontrivial linear manifold of entire functions that is
complete with respect to the norm generated by a given inner product
$\inner{\cdot}{\cdot}_{\cal B}$. We say that ${\cal B}$ is an
{\em (axiomatic) de Branges space} if, for every $f(z)$ in that space,
the following conditions holds:
\begin{enumerate}[({A}1)]
\item For every $w:\im w\neq 0$, the linear functional
        $f(\cdot)\mapsto f(w)$  is continuous;

\item for every non-real zero $w$ of $f(z)$, the function
        $f(z)(z-\cc{w})(z-w)^{-1}$ belongs to ${\cal B}$
        and has the same norm as $f(z)$;

\item the function $f^\#(z):=\cc{f(\cc{z})}$ also belongs to ${\cal B}$
        and has the same norm as $f(z)$.
\end{enumerate}

By the Riesz lemma, (A1) is equivalent to the existence of a
reproducing kernel $k(z,w)$ that belongs to $\cal B$ for every
non-real $w$ and has the property $\inner{k(\cdot,w)}{f(\cdot)}_{\cal
  B}=f(w)$ for every $f(z)\in{\cal B}$. Moreover,
$k(w,w)=\inner{k(\cdot,w)}{k(\cdot,w)}_{\cal B}\ge 0$ where, as a
consequence of (A2), the positivity is strict for every non-real $w$
unless $\cal B\cong\mathbb{C}$; see the proof of Theorem~23 in
\cite{debranges}. Notice that
$k(z,w)=\inner{k(\cdot,z)}{k(\cdot,w)}_{\cal B}$ whenever $z$ and $w$
are both non-real, therefore $k(w,z)=\cc{k(z,w)}$. Finally, due to
(A3) it can be shown that $\cc{k(\cc{z},w)}=k(z,\cc{w})$ for every
non-real $w$; we refer again to the proof of Theorem~23 in
\cite{debranges}.

\begin{remark}
  By construction $k(z,w)$ is entire with respect to its first
  argument and, by (A3), it is anti-entire with respect to the second
  one (once $k(z,w)$, as a function of its second argument, has been
  extended to the whole complex plane \cite[Problem 52]{debranges}).
\end{remark}

There is another way of defining a de Branges space. One starts by
considering an entire function $e(z)$ of the Hermite-Biehler class,
that is, an entire function without zeros in the upper half-plane
$\mathbb{C}^+$ that satisfies the inequality
$\abs{e(z)}>\abs{e^\#(z)}$ for $z\in\mathbb{C}^+$. Then, the {\em
  (canonical) de Branges space} $\mathcal{B}(e)$ associated to $e(z)$
is the linear manifold of all entire functions $f(z)$ such that both
$f(z)/e(z)$ and $f^\#(z)/e(z)$ belong to the Hardy space
$H^2(\mathbb{C}^+)$, and equipped with the inner product
\[
\inner{f(\cdot)}{g(\cdot)}_{\mathcal{B}(e)}:=
\int_{-\infty}^\infty\frac{\cc{f(x)}g(x)}{\abs{e(x)}^2}dx.
\]
$\mathcal{B}(e)$ is indeed a Hilbert space. Both definitions of de
Branges spaces are equivalent in the following sense: Every canonical
de Branges space obeys (A1--A3); conversely, given an axiomatic de
Branges space $\mathcal{B}$ there exists an Hermite-Biehler function
$e(z)$ such that $\mathcal{B}$ coincides with $\mathcal{B}(e)$ as sets
and the respective norms satisfies the equality
$\norm{f(\cdot)}_{\mathcal{B}}=\norm{f(\cdot)}_{\mathcal{B}(e)}$
\cite[Chapter 2]{debranges}. The function $e(z)$ in not unique; a
choice for it is
\begin{equation}\label{eq:e-given-by-k}
e(z)=-i\sqrt{\frac{\pi}{k(w_0,w_0)\im(w_0)}}\left(z-\cc{w_0}\right)k(z,w_0),
\end{equation}
where $w_0$ is some fixed complex number with $\im(w_0)>0$. It is customary
and often useful to decompose $e(z)$ into its real and imaginary parts. Define
\[
a(z):= \frac{e(z)+e^\#(z)}{2},\qquad b(z):= i\frac{e(z)-e^\#(z)}{2}
\]
then $e(z)=a(z)-ib(z)$. Both $a(z)$ and $b(z)$ are real entire functions
in the sense that they obey the identity $f^\#(z)=f(z)$.

The reproducing kernel can be expressed in terms of the function
$e(z)$ (see for instance \cite[Section 5]{kaltenback}) as follows
\[
k(z,w) = \begin{cases}
\frac{e^\#(z)e(\cc{w})-e(z)e^\#(\cc{w})}{2\pi i(z-\cc{w})},&z\ne\cc{w}
\\[2mm]
\frac{i}{2\pi}\left[e^{\#'}(z)e(z)-e'(z)e^\#(z)\right],
	&z=\cc{w}.
\end{cases}
\]
Notice that
\[
k(z,z) = \frac{\abs{e(z)}^2-\abs{e^\#(z)}^2}{4\pi\im(z)} > 0,
	\qquad z\in\mathbb{C}^+,
\]
whereby it follows that $e(z)$ given in (\ref{eq:e-given-by-k}) is indeed a
Hermite-Biehler function. From (\ref{eq:e-given-by-k}) we also conclude
that $k(z,w)$ is always different from zero whenever $z,w\in\mathbb{C}^+$.

An entire function $g(z)$ is said to be associated to a de Branges space
$\mathcal{B}$ if for every $f(z)\in\mathcal{B}$ and $w\in\mathbb{C}$,
\begin{equation*}
\frac{g(z)f(w)-g(w)f(z)}{z-w}\in\mathcal{B}.
\end{equation*}
The set of associated functions is denoted $\assoc\mathcal{B}$.
It is well known that
\begin{equation*}
\label{eq:associated-functions}
\assoc\mathcal{B} = \mathcal{B} + z\mathcal{B};
\end{equation*}
see \cite[Theorem 25]{debranges} and \cite[Lemma 4.5]{kaltenback}
for alternative characterizations. In passing, let us note that
$e(z)\in\assoc\mathcal{B}(e)\setminus\mathcal{B}(e)$; this fact follows easily
from \cite[Theorem 25]{debranges}.
Let us also recall that $\assoc\mathcal{B}(e)$ contains an important family of
entire functions. They are given by
\begin{equation}\label{eq:functions-s}
s_\beta(z):=\frac{i}{2}\left[e^{i\beta}e(z)-e^{-i\beta}e^\#(z)\right]
	= -a(z)\sin\beta + b(z)\cos\beta,\quad \beta\in[0,\pi).
\end{equation}
These real entire functions determine the selfadjoint extensions of the
multiplication operator; see below. Notice that $a(z)=-s_{\pi/2}(z)$ and
$b(z)=s_{0}(z)$. Also, every $s_\beta(z)$ has only real zeros of multiplicity
one and the (sets of) zeros of any pair $s_\beta(z)$ and $s_{\beta'}(z)$ are
always interlaced.

Some of the results obtained in this paper are related to the problem
of telling when a de Branges space contains the function 1. In
connection with this, there is a result \cite[Theorem 1.1]{woracek}),
which was recently generalized in various directions \cite{woracek2},
that characterizes de Branges spaces for which $1$ is an associated
function. This result is of interest to us because it relies on the
distribution of the zero-sets of the functions $s_\beta(z)$.  The
aforementioned theorem may be stated as follows.

\begin{theorem}[Woracek]\label{thm:1-is-assoc-to-B}
  Assume $e(x)\neq 0$ for $x\in\mathbb{R}$ and $e(0)=1$. Let
  $\{x_n\}_{n\in\mathbb{N}}$ be the sequence of zeros of the function
  $s_{\pi/2}(z)$. Also, let $\{x_n^+\}_{n\in\mathbb{N}}$ and
  $\{x_n^-\}_{n\in\mathbb{N}}$ be the sequences of positive,
  respectively negative, zeros of $s_{\pi/2}(z)$, arranged according
  to increasing modulus.  Then an everywhere non-zero, real function
  belongs to $\assoc\mathcal{B}(e)$ if and only if the following
  conditions hold true:
\begin{enumerate}[(C1)]
\item The limit
	$\displaystyle{\lim_{r\to\infty}\sum_{0<|x_n|\le r}
		\frac{1}{x_n}}$
	exists;
\item $\displaystyle{\lim_{n\to\infty}\frac{n}{x_n^{+}}
		=\lim_{n\to\infty}\frac{n}{x_n^{-}}<\infty}$;
\item Assuming that $\{b_n\}_{n\in\mathbb{N}}$ are the zeros of
  $s_\beta(z)$, define
	\[
	h_\beta(z):=\left\{\begin{array}{ll}
			\displaystyle{\lim_{r\to\infty}\prod_{|b_n|\le r}
			\left(1-\frac{z}{b_n}\right)}
				& \text{ if 0 is not a root of } s_\beta(z),
			\\
			\displaystyle{z\lim_{r\to\infty}\prod_{0<|b_n|\le r}
			\left(1-\frac{z}{b_n}\right)}
				& \text{ otherwise. }
			   \end{array}\right.
	\]
	The series
	$\displaystyle{
		\sum_{n\in\mathbb{N}}\abs{\frac{1}
		{x_n^2 h_{0}(x_n)h_{\pi/2}'(x_n)}}}$ is convergent.
\end{enumerate}
Furthermore, $s_{\pi/2}(z)/h_{\pi/2}(z)\in\assoc\mathcal{B}(e)$.
\end{theorem}
\begin{remark}
\begin{enumerate}
\item We warn the reader about the precise meaning of condition (C2), which
	has been formulated under the implicit assumption that
	$\{x_n\}_{n\in\mathbb{N}}$ is not semibounded. If the sequence
	is semibounded, say from below, then (C2) means
	\[
	\lim_{n\to\infty}\frac{n}{x_n^{+}}=0.
	\]
	(The other limit is meaningless.)
      \item The assumption of $e(z)$ having no real zeros amounts to
        no loss of generality (see Lemma~2.4 of \cite{kaltenback3});
        by the same token, it has been assumed that $e(0)=1$. The
        latter was done to ensure that 0 is not a root of
        $s_{\pi/2}(z)=-a(z)$, otherwise the proof of the theorem in
        \cite{woracek} would have been more complicated.

\item Due to the interlacing property mentioned above, $s_\beta(0)=0$ for
	only one value of $\beta$.
\end{enumerate}
\end{remark}

Theorem~\ref{thm:1-is-assoc-to-B} has the following consequence.

\begin{corollary}\label{prop:inverse-problem}
  Let $\{x_n^{(1)}\}_{n\in\mathbb{N}}$ and
  $\{x_n^{(2)}\}_{n\in\mathbb{N}}$ be two unbounded interlaced
  sequences of real numbers, with no finite accumulation points. There
  exists a de Branges space $\cB(e)$ such that these sequences are the
  zero sets of $s_{\pi/2}(z)$ and $s_0(z)$, respectively, and
  $1\in\assoc\cB$ if and only if $\{x_n^{(1)}\}_{n\in\mathbb{N}}$ and
  $\{x_n^{(2)}\}_{n\in\mathbb{N}}$ satisfy
\begin{enumerate}[(C0)]
\item $\min\{x_n^{(2)}\}<\min\{x_n^{(1)}\}$ if the sequences are
  semibounded from below, and $\max\{x_n^{(1)}\}>\max\{x_n^{(2)}\}$ if
  the sequences are semibounded from above,
\end{enumerate}
along with conditions (C1), (C2) and (C3) of
Theorem~\ref{thm:1-is-assoc-to-B}, where in (C3) one substitutes
the zeros of $s_{\pi/2}(z)$ by $\{x_n^{(1)}\}$ and the zeros of
$s_0(z)$ by $\{x_n^{(2)}\}$.
\end{corollary}
\begin{proof}
  In view of Theorem~\ref{thm:1-is-assoc-to-B}, one direction of the
  statement has already been established. Note that by \cite[Chapter
  VII, Theorem 1]{levin} the zero sets of $s_{\pi/2}(z)$ and $s_0(z)$
  should satisfy (C0) when the sequences are semibounded.  In order to
  prove the other direction, we assume w.l.o.g. that the sequences
  have been arranged so that $x^{(2)}_k<x^{(1)}_k<x^{(2)}_{k+1}$.  Let
  us define
\[
e(z) := - h_1(z) -i h_2(z),
\]
where $h_i(z)$ is the function defined in (C3) by means of
$\{x_n^{(i)}\}$. It will suffice to show that $e(z)$ is
Hermite-Biehler which, in our case, reduces to showing that
\[
\im\left[-\frac{h_2(z)}{h_1(z)}\right]> 0,\qquad{z\in\mathbb{C}^+}.
\]
This follows directly from \cite[Chapter VII, Theorem 1]{levin} taking
into account (C0) if the sequences are semibounded. Noting
that $a(z)=-h_1(z)$ and $b(z)=h_2(z)$, and recalling
Theorem~\ref{thm:1-is-assoc-to-B}, the assertion follows.
\end{proof}

\begin{remark}
\begin{enumerate}
\item It is clear that, in Corollary~\ref{prop:inverse-problem},
	the phrase ``$1\in\assoc\cB$'' can be replaced by
	``$\assoc\cB$ contains a zero-free real entire function''.
	See \cite[Lemma 2.4]{kaltenback3}

\item The proof above also tells us that conditions (C1) and (C2) are
	sufficient for the existence of a de Branges space $\cB(e)$, having
	the two sequences as zero sets of the associated functions $s_0(z)$
	and $s_{\pi/2}(z)$. However, we cannot assure in this case that
	$\assoc\cB(e)$ contains a zero-free real entire function.
	Indeed, if (C3) is not satisfied then necessarily $\assoc\cB(e)$
	does not contain such kind of entire functions.

\item Corollary~\ref{prop:inverse-problem} together with the
	results of the next section suggest a method for dealing with
	two-spectra inverse problems. This matter will be treated in a
	forthcoming paper.
\end{enumerate}
\end{remark}

In view of the interlacing property, it seems reasonable to believe
that the zero-sets of essentially any pair $s_\beta(z)$ and
$s_{\beta'}(z)$ bear similar properties as those of $s_0(z)$ and
$s_{\pi/2}(z)$. Thus, one may expect that Theorem~\ref{thm:1-is-assoc-to-B}
can be refined accordingly. This refinement is the aim of
the following results.

\begin{lemma}\label{lemma:first}
Suppose $e(z)$ belongs to the Hermite-Biehler class having no real zeros.
For $\gamma\in(0,\pi)$, let $\breve{e}(z)= - s_\gamma(z) - i s_0(z)$.
Then $\breve{e}(z)$ also belongs to the Hermite-Biehler class and has no
real zeros.
Moreover, $\breve{s}_0(z)=s_0(z)$ and $\breve{s}_{\pi/2}(z)=s_\gamma(z)$.
\end{lemma}
\begin{proof}
A short computation leads to the last assertion.
Also, it is obvious that $\breve{e}(z)$ is not constant and has no real
zeros. By the properties of the functions $s_\beta(z)$, it follows
that the quotient
\[
- \frac{\breve{a}(z)}{\breve{b}(z)}
	= \frac{\breve{s}_0(z)}{\breve{s}_\gamma(z)}
\]
is real and analytic in $\mathbb{C}\setminus\mathbb{R}$. Moreover,
\[
-\im\left[\frac{\breve{a}(z)}{\breve{b}(z)}\right]
	= -(\sin\gamma)\im\left[\frac{a(z)}{b(z)}\right]> 0
\]
for $z\in\mathbb{C}^+$. Thus, by a standard criterion for
Hermite-Biehler functions \cite[Chapter VII]{levin}, the first
assertion follows.
\end{proof}

\begin{lemma}\label{lemma:second}
As sets, $\mathcal{B}(\breve{e})=\mathcal{B}(e)$, therefore
$\assoc\mathcal{B}(\breve{e})=\assoc\mathcal{B}(e)$.
\end{lemma}
\begin{proof}
It suffices to show the two-sided estimate
\[
r_1 \abs{e(x+iy)}\ge \abs{\breve{e}(x+iy)}\ge r_2 \abs{e(x+iy)}
\]
for all $y\ge 0$ and some positive constants $r_1$ and $r_2$.

A short computation leads to the identity
\[
\breve{e}(z)
	= e(z)\left[\frac{1-e^{i(\gamma+\frac{\pi}{2})}}{2}\right]
	\left[1-\frac{1-e^{-i(\gamma-\frac{\pi}{2})}}
	{1-e^{i(\gamma+\frac{\pi}{2})}}\frac{e^\#(z)}{e(z)}\right].
\]
The second factor on the r.h.s.\ is a ($\gamma$-dependent) constant whose
absolute value lies in $(\sqrt{2}/2,1)$ because $\gamma\in(0,\pi)$. Similarly,
\[
\abs{\frac{1-e^{-i(\gamma-\frac{\pi}{2})}}
	{1-e^{i(\gamma+\frac{\pi}{2})}}}\in[0,1).
\]
Finally, $\abs{e(x+iy)}\geq\abs{e^\#(x+iy)}$ whenever $y\ge 0$. The
required inequalities now follows easily.
\end{proof}

\begin{proposition}~\label{thm:1-is-assoc-to-B-extended} Suppose
  $e(x)\neq 0$ for $x\in\mathbb{R}$ and $e(0)=(\sin\gamma)^{-1}$ for
  some fixed $\gamma\in(0,\pi)$. Let $\{x_n\}_{n\in\mathbb{N}}$ be the
  sequence of zeros of the function $s_\gamma(z)$. Also, let
  $\{x_n^+\}_{n\in\mathbb{N}}$ and $\{x_n^-\}_{n\in\mathbb{N}}$ be the
  sequences of positive, respectively negative, zeros of $s_\gamma(z)$,
  arranged according to increasing modulus.  Then a zero-free, real
  entire function belongs to $\assoc\mathcal{B}(e)$ if and only if
  (C1) and (C2) of Theorem~\ref{thm:1-is-assoc-to-B} hold true along
  with the additional condition:
\begin{enumerate}[(C3$\,^\flat$)]
\item For $h_\beta(z)$ defined as in (C3), assume
	$\displaystyle{
		\sum_{n\in\mathbb{N}}\abs{\frac{1}
		{x_n^2 h_0(x_n)h_{\gamma}'(x_n)}}<\infty}$.
\end{enumerate}
Also, $s_\gamma(z)/h_\gamma(z)\in\assoc\mathcal{B}(e)$.
\end{proposition}
\begin{proof}
Combine Lemma~\ref{lemma:first}, Lemma~\ref{lemma:second}
and Theorem~\ref{thm:1-is-assoc-to-B}.
\end{proof}

Obviously, multiplication of a Hermite-Biehler function by a
(non-zero) complex number induces both a unitary transformation
between de Branges spaces and a bijection between the respective sets
of associated functions.

\begin{corollary}
Proposition~\ref{thm:1-is-assoc-to-B-extended} remains valid if one replaces
0 and $\gamma$ by $\beta$ and $\beta'$ such that $0\le\beta<\beta'<\pi$ and
assumes that $e(z)$ has no real zeros.
\end{corollary}
\begin{proof}
Apply Proposition~\ref{thm:1-is-assoc-to-B-extended} to
$\mathring{e}(z)=[e(0)\sin\gamma]^{-1}e(z)$ with $\gamma=\beta'-\beta$.
\end{proof}

\begin{remark}
  The statement of Lemma~\ref{lemma:second} readily generalizes to the
  sets of $N$-associated functions introduced in
  \cite{woracek2}. Consequently, Theorem~3.2 of \cite{woracek2} can be
  accordingly refined (at least for the case of de Branges Hilbert
  spaces). We will not discuss this matter further here; nonetheless,
  see the remark at the end of Section~5.
\end{remark}

The operator of multiplication in $\mathcal{B}$ is defined by
\begin{equation}\label{eq:multiplication-operator}
\dom(S):=\{f(z)\in\mathcal{B}: zf(z)\in\mathcal{B}\},\quad
	(Sf)(z)=zf(z).
\end{equation}
This operator is symmetric, simple, regular and has deficiency indices
$(1,1)$, although it is not necessarily densely defined; see
\cite[Proposition 4.2, Corollary 4.3, Corollary 4.7]{kaltenback}.  The
following characterization of the domain is useful; see
\cite[Theorem~29]{debranges} and \cite[Corollary~6.3]{kaltenback}.
\begin{theorem}
$\cc{\dom(S)}\neq\mathcal{B}$ if and only if there exists $\gamma\in[0,\pi)$
such that $s_\gamma(z)\in\mathcal{B}$. Furthermore,
$\dom(S)^\perp=\Span\{s_\gamma(z)\}$.
\end{theorem}

In this paper we shall deal only with cases where $S$ has domain
dense in $\mathcal{B}$. In passing we note that this assumption
is fulfilled, for instance, when the polynomials are dense in $\cB$.
Conditions for this to happen have been studied in \cite{baranov,kaltenback2}.

There exists an explicit relation between the selfadjoint extensions of $S$
and the entire functions $s_\beta(z)$ defined by (\ref{eq:functions-s}); see
\cite[Propositions 4.6 and 6.1]{kaltenback}:
\begin{proposition}\label{thm:selfadjoint-extension-of-multiplication}
The selfadjoint extensions of $S$ are in one-to-one correspondence with the
set of entire functions $s_\beta(z)$, $\beta\in[0,\pi)$. They are
given by
\begin{gather}
\dom(S_\beta) =
	\left\{g(z)=\frac{f(z)-\frac{s_\beta(z)}{s_\beta(z_0)}f(z_0)}{z-z_0},
	\quad f(z)\in\mathcal{B},\quad z_0:s_\beta(z_0)\neq 0\right\},
	\label{eq:non-standard}
\\[2mm]
(S_\beta g)(z) = z g(z) + \frac{s_\beta(z)}{s_\beta(z_0)}f(z_0).\nonumber
\end{gather}
Moreover, $\Sp(S_\beta)=\left\{x_n^{(\beta)}\in\mathbb{R}:
s_\beta(x_n^{(\beta)})=0\right\}$. The associated eigenfunctions (up to
normalization) are $g_n^{(\beta)}(z):=\frac{s_\beta(z)}{z-x_n^{(\beta)}}$.
\end{proposition}

\begin{remark}
The derivation of the expression for the selfadjoint extensions discussed
in \cite{kaltenback} is done in the broader context of selfadjoint relations
(so it is valid even when the codimension of $\cc{\dom(S)}$ is one).
However, it might be beneficial to see the connection between the
characterization of selfadjoint extensions given by
Proposition~\ref{thm:selfadjoint-extension-of-multiplication} and the standard
way {\em \`a la} von Neumann. We first note that
\[
n(z,w_0):=
	\frac{k(z,w_0)}{\sqrt{k(w_0,w_0)}}\in\Ker\left(S^*-\cc{w_0}I\right),
	\quad \im(w_0)> 0.
\]
The choice on the phase of $w_0$ is related to the one made on
(\ref{eq:e-given-by-k}) so it is not an actual restriction.
Also, $\norm{n(\cdot,w_0)}=1$. Since $S$ has deficiency indices $(1,1)$, by
the standard theory its selfadjoint extensions are given by
\begin{gather*}
\dom(S_\beta) =
	\left\{g(z) = h(z) + pe^{-i\beta}n(z,\cc{w_0}) + pe^{i\beta}n(z,w_0),
	\,\, h(z)\in\dom(S),\,\, p\in\mathbb{C}\right\},
\\[2mm]
(S_\beta g)(z) = z h(z)
	+ pe^{-i\beta}w_0 n(z,\cc{w_0}) + pe^{i\beta}\cc{w_0} n(z,w_0),
\end{gather*}
where $\beta\in[0,\pi)$. Due to (A3), $n(z,\cc{w_0})=\cc{n(\cc{z},w_0)}$,
so, by (\ref{eq:e-given-by-k}), doing some rearrangement and
then using (\ref{eq:functions-s}), we obtain
\begin{gather}
\dom(S_\beta) =
	\left\{g(z)\!=\!h(z) - ire^{-i\beta}\frac{e^\#(z)}{z-w_0}
			+ ire^{i\beta}\frac{e(z)}{z-\cc{w_0}},
	\,\, h(z)\in\dom(S),\,\, r\in\mathbb{C}\right\},
	\label{eq:von-newmann}
\\[2mm]
(S_\beta g)(z)
	= z h(z)
		- ire^{-i\beta}w_0\frac{e^\#(z)}{z-w_0}
		+ ire^{i\beta}\cc{w_0}\frac{e(z)}{z-\cc{w_0}}
	= z g(z) - 2 r s_\beta(z).\nonumber
\end{gather}
Finally, define
\[
f(z) := (z-w_0)h(z) + 2 r \im(w_0)e^{i\beta}\frac{e(z)}{z-\cc{w_0}}.
\]
The function $f(z)$ is entire as long as either is $h(z)$; notice that
$e(\cc{w_0})=0$ so the last term above is entire. It is not difficult to verify
that $f(z)\in\mathcal{B}(e)$ if and only if $h(z)\in\dom(S)$. Moreover,
$f(w_0)=-2rs_\beta(w_0)$. We obtain (\ref{eq:non-standard}) after
substituting $h(z)$ by $f(z)$ in (\ref{eq:von-newmann}).

In passing, and for the sake of completeness, we note that a computation in
a similar vein yields the following description of the adjoint operator:
\begin{gather*}
\dom(S^*) =
	\left\{g(z)
		 \!=\!\frac{f(z)-\frac{e(z)}{e(z_0)}f(z_0)}{z-z_0} +
		 \frac{h(z)-\frac{e^\#(z)}{e^\#(\cc{z_0})}
						h(\cc{z_0})}{z-\cc{z_0}}:
	f(z),h(z)\in\mathcal{B},\ \im(z_0)>0\right\}\!,
\\[2mm]
(S^*g)(z) = z g(z) + \frac{e(z)}{e(z_0)}f(z_0)
		+ \frac{e^\#(z)}{e^\#(\cc{z_0})}h(\cc{z_0}).
\end{gather*}
\end{remark}

For the selfadjoint extension $S_\beta$ of $S$, the spectral measure is
given by
\begin{equation}
\label{eq:spectral-measure}
m_\beta(x) =
	\sum_{\stackrel{x_n\in\Sp(S_\beta)}{x_n\le x}}
	\norm{k(\cdot,x_n)}^{-2}.
\end{equation}
These spectral measures allow us
to compute the inner product of $\cB$ in different ways.
We note that $\{k(z,x_n)\norm{k(\cdot,x_n)}^{-1}\}$ is an orthonormal basis
of $\cB$ when $\{x_n\}=\Sp(S_\beta)$ thus one readily obtain
\begin{equation}
\label{eq:equivalence-inner}
\inner{f(\cdot)}{g(\cdot)}_{\mathcal{B}}=
\int_{-\infty}^\infty\cc{f(x)}g(x)dm_\beta(x).
\end{equation}
Notice that
\begin{equation}\label{eq:spectral-weights}
\norm{k(\cdot,x_n)}^2 =
k(x_n,x_n)=-\frac{1}{\pi}s_{\gamma+\frac{\pi}{2}}(x_n)s'_{\gamma}(x_n),
\end{equation}
for $x_n\in\Sp(S_\gamma)$ and
$\gamma,\gamma+\frac{\pi}{2}\in\mathbb{R}/[0,\pi)$,
where the second equality comes from a standard expression for $k(z,w)$
\cite[Equation 2.1]{kaltenback3}.
\begin{remark}
\begin{enumerate}
\item A straightforward consequence of (\ref{eq:equivalence-inner}) is
  that a function in $\cB$ restricted to the real line is in
  $L^2(m_\beta)$. Moreover, from the simplicity of $S_\beta$ it
  follows that $\cB$ fills $L^2(m_\beta)$ (see \cite[Problem
  69]{debranges}) in the sense that, given $\varphi(x)\in
	L^2(m_\beta)$, there exists $f(z)\in\cB$ such that
	$\norm{\varphi(\cdot)-f(\cdot)}_{L^2(m_\beta)}=0$.
\item Also, by \cite[Problem 69]{debranges}, if $g(z)\in\assoc\cB$
	satisfies $\norm{g(\cdot)}_{L^2(m_\beta)}=0$ then $g(z)$ is a
	constant multiple of $s_\beta(z)$ (the converse statement is
	obvious).
\end{enumerate}
\end{remark}

The following result is analogous to
Theorem~\ref{thm:1-is-assoc-to-B}, although it provides necessary and
sufficient conditions for a subclass of the spaces considered in that
theorem.
\begin{proposition}\label{prop:1-in-dB-boosted}
Let $\cB(e)$ be a de Branges space such that $S$ is densely defined.
Then a real entire zero-free function lies in $\cB(e)$ if and only if
condition (C1) and (C2) of
Theorem~\ref{thm:1-is-assoc-to-B} hold along with
\begin{enumerate}[(C3$\,^\sharp$)]
\item $\displaystyle{\sum_{x_n\in\Sp(S_\gamma)}\abs{
\frac{1}{h_{0}(x_n)h'_\gamma(x_n)}}<\infty,\quad \gamma\in(0,\pi)}$,
\end{enumerate}
where $h_\beta(z)$ is defined in (C3) of the theorem already mentioned.
\end{proposition}
\begin{proof}
  We will prove the proposition for $\gamma=\pi/2$; by an
  argumentation like the one used in Lemma~\ref{lemma:first}, the
  proof extends to arbitrary $\gamma$.

  Suppose that a real entire zero-free function $g(z)$ is in $\cB(e)$.
  Then $1\in\cB(\breve{e})$, where $\breve{e}(z)=e(z)/g(z)$. Consider
  the operator of multiplication $\breve{S}$ in $\cB(\breve{e})$.
  Clearly, $\breve{S}$ is densely defined together with $S$, so one
  has
\begin{equation}\label{eq:leading-to-our-C3}
  \norm{1}_{\cB(\breve{e})}^2 =
  -\pi\sum_{x_n\in\Sp(S_\frac{\pi}{2})}
  \frac{1}{\breve{s}_{0}(x_n)\breve{s}'_{\frac{\pi}{2}}(x_n)}<\infty,
\end{equation}
where (\ref{eq:spectral-weights}) has been used together with the fact
that $\Sp(S_\gamma)=\Sp(\breve{S}_\gamma)$.  Since
$1\in\assoc\cB(\breve{e})$, then $\breve{s}_{\frac{\pi}{2}}(z)$ is
of bounded type in the upper half-plane by
\cite[Lemma~2.1]{woracek}. Thus, from \cite[Proposition~2.2]{woracek}
we obtain that (C1), (C2), and the identities
\begin{equation}\label{eq:s-h}
\breve{s}_{0}(z)=ch_0(z),\qquad
\breve{s}_{\frac{\pi}{2}}(z)=-h_{\frac{\pi}{2}}(z),
\end{equation}
with $c>0$, hold. These expressions for $\breve{s}_{0}$ and
$\breve{s}_{\frac{\pi}{2}}$, in conjunction with
(\ref{eq:leading-to-our-C3}), imply (C3$\,^\sharp$).

Note that Proposition~2.2 of \cite{woracek} is based on
\cite[Theorem~3]{krein3-5} (see also \cite[Theorem~6.17]{rosenblum})
and \cite[Chapter 5 Theorem~11]{levin}, and, by the latter, one would
have obtained a factor $e^{ikz}$ ($k\in\mathbb{R}$) on the r.h.s of
each equality in (\ref{eq:s-h}). Due to the fact that the entire
functions $\breve{s}_{0}(z)$ and $\breve{s}_{\frac{\pi}{2}}(z)$ are
real, $k$ must be zero.

Now, assume that (C1), (C2), and (C3$\,^\sharp$) are fulfilled. Then,
$h_0(z)$ and $h_{\frac{\pi}{2}}(z)$ are well defined. Set
\begin{equation*}
  \tilde{e}(z) := -h_{\frac{\pi}{2}}(z)-i\tilde{c}h_0(z)
\end{equation*}
for $\tilde{c}>0$ to be chosen later. As in the proof of
Corollary~\ref{prop:inverse-problem}, one establishes that
$\tilde{e}(z)$ is Hermite-Biehler. By an appropriate choice of
$\tilde{c}$, this Hermite-Biehler function and the function
$e(z)=-s_{\frac{\pi}{2}}(z)-is_0(z)$ satisfies
$\tilde{e}(z)=j(z)e(z)$, where $j(z)$ is a real entire and zero-free
function. Hence, in view of Lemma~2.4 of \cite{kaltenback3}, the
operator of multiplication $\tilde{S}$ is densely defined in
$\cB(\tilde{e})$, thus
$\tilde{s}_\frac{\pi}{2}(z)=h_{\frac{\pi}{2}}(z)\in\assoc\cB(\tilde{e})
\setminus\cB(\tilde{e})$.


On the basis of (\ref{eq:spectral-measure}), (\ref{eq:equivalence-inner}),
(\ref{eq:spectral-weights}) and condition (C3$\,^\sharp$), it follows that
\begin{equation*}
\norm{1}_{L^2(\tilde{m}_\frac{\pi}{2})}^2 =
	-\frac{\pi}{\tilde{c}}\sum_{x_n\in\Sp(S_\frac{\pi}{2})}
	\frac{1}{h_{0}(x_n)h'_{\frac{\pi}{2}}(x_n)}<\infty.
\end{equation*}
By the first item of the last remark above, there exists
$f(z)\in\cB(\tilde{e})$ such that the
$L^2(\tilde{m}_\frac{\pi}{2})$-norm of $g(z):=1-f(z)$ equals zero.  By
the second item of the same remark, and noting that necessarily
$g(z)\in\assoc\cB(\tilde{e})$, it turns out that
$g(z)=w\tilde{s}_\frac{\pi}{2}(z)$ for some $w\in\mathbb{C}$.  Recalling 
(\ref{eq:functions-s}), a computation yields
\[
\frac{1+w\tilde{s}_{\frac{\pi}{2}}(x)}{\tilde{e}(x)}
	= \frac{1}{\abs{\tilde{e}(x)}} - w\cos\varphi(x),\qquad
	x\in\mathbb{R},
\]
where $\varphi(x)$ is the phase function associated to $\tilde{e}(z)$
given by the identity $\tilde{e}(x)=
e^{-i\varphi(x)}\abs{\tilde{e}(x)}$, $x\in\mathbb{R}$. By \cite[Problem
48]{debranges}, $\varphi(x)$ is a continuous (indeed, differentiable)
strictly increasing function; note that the zeros of
$\tilde{s}_{\frac{\pi}{2}}(z)$ satisfies the identity
$\varphi(x_n)=\pi/2 \mod\pi$.  Since $f(z)\in\cB(\tilde{e})$, it
follows that
\begin{equation}\label{eq:osc-limit}
\frac{1}{\abs{\tilde{e}(x)}} - w\cos\varphi(x)\to 0
\text{ as } \abs{x}\to\infty.
\end{equation}
This implies that necessarily $w\in\mathbb{R}$. Furthermore, since
the zero-set of $\tilde{s}_\beta(z)$ is unbounded, $\varphi(x)$ is
unbounded so there exists an unbounded set
$\{y_n\}_{n\in\mathbb{N}}$ such that, for all $n\in\mathbb{N}$,
\[
w\cos\varphi(y_n) = -\abs{w}.
\]
But this condition collides with (\ref{eq:osc-limit}) unless $w=0$. We
therefore conclude that $f(z)\equiv 1$, in other words, $1\in\cB(\tilde{e})$.
Finally, recalling the relation between $\tilde{e}(z)$ and
$e(z)$, we obtain that $j(z)^{-1}\in\cB(e)$.
\end{proof}

\begin{remark}
  After we submitted the present paper for publication, we learned of
  \cite{woracek2} which gives necessary and sufficient conditions for
  a real zero-free function to be $N$-associated to a de Branges
  Pontryagin space. This result contains as a particular case
  Theorem~\ref{thm:1-is-assoc-to-B} and, in conjunction with
  Lemma~\ref{lemma:first}, it also contains
  Proposition~\ref{prop:1-in-dB-boosted}. Our alternative approach to
  this proposition provides a simpler proof than the one of the more
  general case treated in \cite{woracek2}, in part due to the fact
  that we are dealing with Hilbert spaces (rather than Pontryagin
  spaces), but also because of the techniques of spectral theory we used.
\end{remark}

Concluding this section, we note that if one substitutes (C3) by
(C3$\,^\sharp$) in Corollary~\ref{prop:inverse-problem}, one obtains
necessary and sufficient conditions for two sequences to generate a de
Branges space containing a real entire zero-free function.

\section{Representation by symmetric and entire operators}
It is known that every entire operator $A$ generates a representation
of $\cH$ as a de Branges space in which $A$ becomes the operator of
multiplication by the independent variable \cite[Chapter 2]{gorbachuk}. 
This assertion is in fact true for any operator in 
$\text{Sym}^{(1,1)}_\text{R}({\cH})$. Moreover, recent results 
\cite{martin} show that for every
regular, closed, simple symmetric operator $A$ (not necessarily
densely defined) with deficiency indices $(1,1)$, there exists a de
Branges space such that the operator of multiplication in it 
is unitarily equivalent to $A$.

The representation proposed here not only provides an alternative
approach, for the case of densely defined operators, to the theory
developed in \cite{martin} but also allows to construct explicitly 
the isometry that associates every element of $\cH$ with the
corresponding entire function in the de Branges space. Furthermore,
this representation reduces to the one derived from Krein's work for
the case of entire operators and, for that reason, it is suitable for the
purpose of the present work. It is also worth remarking that our
representation makes possible the treatment of generalized entire
operators in a straightforward way (see Section 5).

 The representation proposed here is realized by associating to
the operator $A$ a certain nowhere-zero, vector-valued entire function
\[
\mathbb{C}\ni z\mapsto\xi(z)\in\Ker(A^*-zI).
\]
Fix an operator $A\in\text{Sym}^{(1,1)}_\text{R}({\cal H})$ and one of
its selfadjoint extensions $A_\gamma$, with $\gamma\in[0,\pi)$.
Let $g^{(\gamma)}(z)$ be a real
entire function such that its zero-set coincides with $\Sp(A_\gamma)$.  Since
$\Sp(A_\gamma)$ has no finite accumulation points, there always exists
such an entire function.

Consider the function
$\psi^{(\gamma)}(z)$ introduced at the beginning of the Section~2
(see (\ref{eq:cayley})). By
\cite[Section 2, Theorem 7.1]{gorbachuk}, there is a complex conjugation $C$
that commutes with $A$, hence with all its selfadjoint extensions.
Therefore,
\begin{equation}
  \label{property-C}
  C\psi^{(\gamma)}(z)=\psi^{(\gamma)}(\cc{z})\,.
\end{equation}
Define
\begin{equation}\label{eq:xi}
  \xi^{(\gamma)}(z):= g^{(\gamma)}(z)\psi^{(\gamma)}(z).
\end{equation}
\begin{lemma}\label{lemma:c-on-g}
$\xi^{(\gamma)}(z)$ is entire, everywhere non-zero and
has image in $\Ker(A^*-zI)$ for every $z\in\mathbb{C}$. Moreover,
\begin{equation*}
  C\xi^{(\gamma)}(z)=\xi^{(\gamma)}(\cc{z})\,.
\end{equation*}
\end{lemma}
\begin{proof}
The function $\xi^{(\gamma)}(z)$ is entire and everywhere non-zero
by rather obvious reasons. Also, we already know that
$\xi^{(\gamma)}(z)\in\Ker(A^*-zI)$ for $z\not\in\Sp(A_\gamma)$ so
we must verify the assertion only for $z=x_n\in\Sp(A_\gamma)$. In a
neighborhood of $x_n$,
\[
\psi^{(\gamma)}(z) =
	(z-x_n)^{-1}\kappa_n +
	\text{a function analytic at }x_n,
\]
where the residue $\kappa_n$ can be shown to lie in $\Ker(A^*-x_nI)$
(see the proof of Proposition 2 in \cite{us}).

The second part of the assertion directly follows from
(\ref{property-C}) and the fact that $g^{(\gamma)}(z)$ is real.
\end{proof}

\begin{remark}
Note that, although the zero-set of $g^{(\gamma)}(z)$ is determined
by the choice of the selfadjoint extension, one has still the freedom
of multiplying it by an arbitrary zero-free, real entire function. This
implies that $\xi^{(\gamma')}(z)=j(z)\xi^{(\gamma)}(z)$, where $j(z)$
is a zero-free real entire function. That is, up to multiplication by a
zero-free real entire function, $\xi^{(\gamma)}(z)$ does not depends on
the parameter $\gamma$. In view of this, the function
$\xi^{(\gamma)}(z)$ will be denoted just as $\xi(z)$.
\end{remark}

Now define
\begin{equation}\label{eq:defining-phi}
\left(\Phi\varphi\right)(z):=\inner{\xi(\cc{z})}{\varphi},\qquad
	\varphi\in\cH.
\end{equation}
$\Phi$ maps $\cH$ onto a certain linear manifold $\widehat{\cH}$ of
entire functions. Since $A$ is simple, it follows that $\Phi$ is
injective.  A generic element of $\widehat{\cH}$ will be often denoted
by $\widehat{\varphi}(z)$, as a reminder of the fact that it is the
image under $\Phi$ of a unique element $\varphi\in\cH$. 

The linear space $\widehat{\cH}$ is turned into a Hilbert space by defining
\begin{equation*}
  \inner{\widehat{\eta}(\cdot)}{\widehat{\varphi}(\cdot)}:=
\inner{\eta}{\varphi}\,.
\end{equation*}
Clearly, $\Phi$ is an isometry from $\cH$ onto $\widehat{\cH}$.
\begin{proposition}
$\widehat{\cH}$ is an axiomatic de Branges space.
\end{proposition}
\begin{proof}
We shall verify (A1--A3) in that order.

1) Define $k(z,w):=\inner{\xi(\cc{z})}{\xi(\cc{w})}$. Then, for every
$\widehat{\varphi}\in\widehat{\cH}$,
\[
\inner{k(\cdot,w)}{\widehat{\varphi}(\cdot)}
	=\inner{\xi(\cc{w})}{\varphi}
	= \widehat{\varphi}(w).
\]

2) Assume now that $\widehat{\varphi}(z)$ has a non-real zero
$w\in\mathbb{C}$. That is, $\inner{\xi(\cc{w})}{\varphi}=0$ which implies
that $\varphi\in\ran(A-wI)$. Thus, it makes sense to set
$\eta=(A-\cc{w}I)(A-wI)^{-1}\varphi$, with image
$\widehat{\eta}(z)\in\widehat{\cH}$ under $\Phi$. Furthermore, a rather
straightforward computation \cite{us} shows that
\[
\widehat{\eta}(z) = \frac{z-\cc{w}}{z-w}\widehat{\varphi}(z).
\]
Since $\eta$ and $\varphi$ are related by a Cayley transform, it follows that
$\norm{\widehat{\eta}(\cdot)}=\norm{\widehat{\varphi}(\cdot)}$.

3) Given any $\widehat{\varphi}(z)\in\widehat{\cH}$, consider the
entire function
\[
\widehat{\varphi}^\#(z):=\cc{\widehat{\varphi}(\cc{z})}.
\]
Since by Lemma~\ref{lemma:c-on-g} we have
\[
\cc{\widehat{\varphi}(\cc{z})}
	= \cc{\inner{\xi(z)}{\varphi}}
	= \inner{C\xi(z)}{C\varphi}
	= \inner{\xi(\cc{z})}{C\varphi},
\]
it follows that $\widehat{\varphi}^\#(z)$ also belongs to $\widehat{\cH}$ and
has the same norm as $\widehat{\varphi}(z)$.
\end{proof}
The reproducing kernel in $\widehat{\cH}$ is given by
\begin{equation*}
k(z,w) = g^{(\gamma)}(z)g^{(\gamma)}(\cc{w})
		\inner{\psi^{(\gamma)}(\cc{z})}{\psi^{(\gamma)}(\cc{w})}.
\end{equation*}
Although the choice of $z_0$ and $\psi_0\in\Ker(A^*-z_0I)$ that enters
in the definition of $\psi^{(\gamma)}(z)$ (see (\ref{eq:cayley}))
is arbitrary, w.l.o.g. we can conveniently assume that $\im(z_0)>0$ and
$\norm{\psi_0}=1$. In that case, a computation yields
\begin{equation*}
k(z,z_0) = g^{(\gamma)}(z)g^{(\gamma)}(\cc{z_0})
	\left[1 + (z-z_0)\inner{\psi_0}{(A_\gamma-zI)^{-1}\psi_0}\right].
\end{equation*}
Therefore, recalling (\ref{eq:e-given-by-k}),
\begin{equation*}
e(z) = -i\left[\frac{\pi}{\im(z_0)}\right]^{\frac12}
	\frac{g^{(\gamma)}(\cc{z_0})}{|g^{(\gamma)}(\cc{z_0})|}
	g^{(\gamma)}(z)(z-\cc{z_0})
	\left[1 + (z-z_0)\inner{\psi_0}{(A_\gamma-zI)^{-1}\psi_0}\right].
\end{equation*}

It is known that the growth of $\cB(e)$ coincides with the growth of $e(z)$;
see \cite{kaltenback3}.
The last expression shows that the growth of $e(z)$, hence of $\widehat{\cH}$,
is governed by the growth of $g^{(\gamma)}(z)$.

The next statement can easily be proved
(cf.\ \cite[Chapter~1, Theorem~2.2]{gorbachuk}). We leave the details to the
reader.

\begin{proposition} Let $S$ be the multiplication operator on
  $\widehat{\cH}$ given by (\ref{eq:multiplication-operator}). Then,
\begin{enumerate}
\item $S=\Phi A\Phi^{-1}$ and $\dom(S):=\Phi\dom(A)$ (thus
	$S$ is densely defined);
\item the selfadjoint extensions of $S$ are in one-one correspondence
	with the selfadjoint extensions of $A$.
\end{enumerate}
\end{proposition}

The following assertions give explicit characterizations of when an
operator is entire. We emphasize that
Theorem~\ref{cor:spectrum-tells-if-operator-is-entire}
rests entirely upon conditions on the spectra of selfadjoint extensions.

\begin{proposition}
\label{prop:when-the-operator-is-entire}
$A\in\ourclass$ is entire if and only if $\widehat{\cH}$ contains a
real zero-free entire function.
\end{proposition}

\begin{proof}
Let $g(z)\in\widehat{\cH}$ be the function whose
existence is assumed. Clearly there exists (a unique) $\mu\in\cH$ such
that $g(z)\equiv\inner{\xi_1(\cc{z})}{\mu}$. Therefore, $\mu$ is never
orthogonal to $\Ker(A^*-zI)$ for all $z\in\mathbb{C}$.
That is, $\mu$ is an entire gauge for the operator $A$.

The necessity is established by noting that the image of the entire
gauge under $\Phi$ is a zero-free function.
\end{proof}

\begin{theorem}
\label{cor:spectrum-tells-if-operator-is-entire}
For $A\in\ourclass$, consider the selfadjoint extensions $A_0$ and
$A_{\gamma}$, with $0<\gamma<\pi$. Then
$A$ is entire if and only if $\Sp(A_0)$ and $\Sp(A_{\gamma})$ obey
conditions (C1), (C2) and (C3$\,^\sharp$) of
Proposition~\ref{prop:1-in-dB-boosted}.
\end{theorem}
\begin{proof}
Apply Proposition~\ref{prop:1-in-dB-boosted} along with
Proposition~\ref{prop:when-the-operator-is-entire}.
\end{proof}

\begin{proposition}
\label{thm:krein-particular-case}
Assume $1\in\widehat{\cH}$. Then there exists $\mu\in\cH$
such that
\[
g^{(\gamma)}(z)=\frac{1}{\inner{\psi^{(\gamma)}(\cc{z})}{\mu}}
\]
and $C\mu=\mu$. Moreover, $\mu$ is the unique entire gauge of $A$ modulo
a real scalar factor.
\end{proposition}
\begin{proof}
Necessarily, $1\equiv\inner{\xi(\cc{z})}{\mu}$
for some $\mu\in\cH$. By (\ref{eq:xi}), and taking into account the
occurrence of $C$, one obtains the stated expression for $g^{(\gamma)}(z)$.
By the same token, the reality of $\mu$ is shown.

Suppose that there are two real entire gauges $\mu$ and $\mu'$.
The discussion in Paragraph~5.2 of \cite{gorbachuk} shows that
$(\Phi_{\mu}\mu')(z)=ae^{ibz}$ with $a\in\mathbb{C}$ and $b\in\mathbb{R}$.
Due to the assumed reality, one concludes that $b=0$ and $a\in\mathbb{R}$.
\end{proof}

\begin{remark}
Proposition~\ref{thm:krein-particular-case} shows that Krein's theory of
representation by entire operators is a particular case of the
representation proposed here.
\end{remark}

\section{Operators entire in the generalized sense}

In this section we give a spectral characterization of operators
in $\ourclass$ that are entire with respect to a generalized gauge
\cite[Chapter 3, Section 9]{gorbachuk}. This section was added as a
result of a suggestion of the reviewer.

Given $A\in\ourclass$, let $\cH_{+}$ be the set $\dom(A^*)$ equipped
with the graph norm
\begin{equation*}
\norm{\varphi}_{+}^2:= \norm{\varphi}^2 + \norm{A^*\varphi}^2,\qquad
	\varphi\in\dom(A^*).
\end{equation*}
Let $\cH_{-}$ be the completion of $\cH$ under the norm
\begin{equation*}
\norm{\eta}_{-}:=
	\sup_{\varphi\in\cH_{+}}
	\frac{\abs{\inner{\eta}{\varphi}}}{\norm{\varphi}_{+}},\qquad
	\eta\in\cH;
\end{equation*}
the elements of $\cH_{-}$ are the continuous linear functionals on $\cH_{+}$.
In this way one obtains the scale of Hilbert spaces
$\cH_{+}\subset\cH\subset\cH_{-}$ associated to $A^*$, where the embeddings
are dense and continuous. Sticking to the standard notation,
for $\eta\in\cH_{-}$ and $\varphi\in\cH_{+}$ we define
$\inner{\eta}{\varphi}:=\eta(\varphi)$ so accordingly
$\inner{\varphi}{\eta}:=\cc{\eta(\varphi)}$.

Given a selfadjoint extension $A_\beta$ of $A$ and $z\not\in\Sp(A_\beta)$,
let $R_z^{(\beta)}$ be the extension of $(A_\beta-zI)^{-1}$ from $\cH$ to
$\cH_{-}$. This operator satisfies the identity
\begin{equation*}
\inner{R_z^{(\beta)}\eta}{\varphi} =
	\inner{\eta}{(A_\beta-\cc{z}I)^{-1}\varphi},\qquad
	\eta\in\cH_{-},\qquad\varphi\in\cH.
\end{equation*}
It is straightforward to verify that $R_z^{(\beta)}$ maps $\cH_{-}$ into
$\cH$. It also satisfies the extended resolvent identity
\begin{equation*}
R_z^{(\beta)} - R_w^{(\beta)}
	= (z-w)(A_\beta-zI)^{-1}R_w^{(\beta)}
	= (z-w)(A_\beta-wI)^{-1}R_z^{(\beta)}.
\end{equation*}

A complex conjugation on $\cH$ is extended to $\cH_{-}$ by defining
\begin{equation*}
\inner{C\eta}{\varphi}:=\cc{\inner{\eta}{C\varphi}},\qquad
	\eta\in\cH_{-},\qquad \varphi\in\cH_{+}.
\end{equation*}
We say that $\eta\in\cH_{-}$ is real if $C\eta=\eta$.

Let $\xi(z)$ be the entire vector-valued function defined by (\ref{eq:xi}).
Let us recall that $\xi(z)\in\Ker(A^*-zI)\subset\dom(A^*)$, therefore
the linear map $\Phi$ defined by (\ref{eq:defining-phi}) on $\cH$ can be
extended to $\cH_{-}$ in the obvious manner.

\begin{proposition}
\label{prop:assoc-of-hat-h}
$\displaystyle{
\assoc\widehat\cH
	= \Phi\cH_{-}
	:= \left\{\widehat{\eta}(z)=\inner{\xi(\cc{z})}{\eta}:
		\eta\in\cH_{-}\right\}.}$
\end{proposition}
\begin{proof}
Given some arbitrary $w\in\mathbb{C}$, choose a selfadjoint
extension $A_\beta$ such that $w\not\in\Sp(A_\beta)$. Let us recall that 
\[
\xi(z) = g^{(\beta)}(z)\psi^{(\beta)}(z),
\]
where $\psi^{(\beta)}(z)$ is given by (\ref{eq:cayley}). By assumption
$g^{(\beta)}(w)\ne 0$ (and also $g^{(\beta)}(\cc{w})\ne 0$ ---this fact is 
used below). A computation involving the first resolvent 
identity yields the equality
\begin{equation}\label{eq:tool}
\frac{\xi(z)-\xi(w)}{z-w} =
	(A_\beta-wI)^{-1}\xi(z) +
	\frac{g^{(\beta)}(z)-g^{(\beta)}(w)}{(z-w)g^{(\beta)}(w)}\xi(w). 
\end{equation}

Now consider $\widehat{\eta}(z)=\inner{\xi(\cc{z})}{\eta}$ with 
$\eta\in\cH_{-}$ and $\widehat{\varphi}(z)=\inner{\xi(\cc{z})}{\varphi}$ 
with $\varphi\in\cH$. Then 
$\widehat{\eta}(w)\varphi-\widehat{\varphi}(w)\eta\in\cH_{-}$. Moreover, 
\begin{align*}
\frac{\widehat{\eta}(w)\widehat{\varphi}(z) -
		 \widehat{\eta}(z)\widehat{\varphi}(w)}{z-w}
	&= \frac{1}{z-w}\inner{\xi(\cc{z})}
		{\widehat{\eta}(w)\varphi-\widehat{\varphi}(w)\eta}
	\\[1mm]
	&= \inner{\frac{\xi(\cc{z})-\xi(\cc{w})}{\cc{z}-\cc{w}}}
		{\widehat{\eta}(w)\varphi-\widehat{\varphi}(w)\eta}
	\\[1mm]
	&= \inner{(A_\beta-\cc{w})^{-1}\xi(\cc{z})}
		{\widehat{\eta}(w)\varphi-\widehat{\varphi}(w)\eta}
	 = \inner{\xi(\cc{z})}{\tau},
\end{align*}
where $\tau := R_w^{(\beta)}
	\left[\widehat{\eta}(w)\varphi-\widehat{\varphi}(w)\eta\right]\in\cH.$
In the last computation we have used (\ref{eq:tool}) and the fact that
$\inner{\xi(\cc{w})}{\widehat{\eta}(w)\varphi-\widehat{\varphi}(w)\eta}=0$.
It follows that $\Phi\cH_{-}\subset\assoc\widehat{\cH}$.

Next, consider $g(z)\in\assoc\widehat{\cH}$. Then there exist two functions
$f(z),h(z)\in\widehat{\cH}$ such that $g(z)=f(z)+zh(z)$. Since the
multiplication operator $S$ is densely defined, there exists a sequence
$\{h_n(z)\}_{n\in\mathbb{N}}\subset\dom(S)$ that is $\widehat{\cH}$-norm
convergent to $h(z)$. Moreover, $f(z)+zh_n(z)$ converges to $g(z)$
uniformly on compact subsets. For every $n\in\mathbb{N}$ we have
$h_n(z)=\inner{\xi(\cc{z})}{\eta_n}$ for some unique
$\eta_n\in\dom(A)$, and also $f(z)=\inner{\xi(\cc{z})}{\psi}$ for
$\psi\in\cH$. Set $\tau_n=\psi + A\eta_n$. Since
\begin{equation*}
\norm{\tau_m-\tau_n}_{-}
	=   \sup_{\varphi\in\cH_{+}}
		\frac{\abs{\inner{\left(\eta_m-\eta_n\right)}
			{A^*\varphi}}}{\norm{\varphi}_{+}}
	\le \sup_{\varphi\in\cH_{+}}
		\frac{\norm{\eta_m-\eta_n}\norm{A^*\varphi}}
			{\norm{\varphi}_{+}}
	\le \norm{\eta_m-\eta_n},
\end{equation*}
it follows that the sequence $\{\tau_n\}_{n\in\mathbb{N}}$ converges to some
$\tau\in\cH_{-}$ which in turn satisfies
$g(z)=\inner{\xi(\cc{z})}{\tau}$. Therefore,
$\assoc\widehat{\cH}\subset\Phi\cH_{-}$.
\end{proof}

\begin{definition}
An operator $A\in\ourclass$ is said to be {\em entire with respect
to a generalized gauge}, or just {\em entire in the generalized sense},
if there exists $\mu\in\cH_{-}$ such that $\inner{\xi(\cc{z})}{\mu}\neq 0$
for all $z\in\mathbb{C}$.
\end{definition}

Note that this definition becomes equivalent to
Definition~\ref{def:entire-operator} when the linear functional $\mu$ can
be identified with an element in $\cH$.
The following assertion is an obvious consequence of
Proposition~\ref{prop:assoc-of-hat-h}.

\begin{proposition}
An operator $A\in\ourclass$ is entire in the generalized sense if and
only if $\assoc\widehat{\cH}$ contains a real zero-free entire function.
\end{proposition}

Finally, we have:

\begin{theorem}\label{thm: a-is-entire-generalized}
For $A\in\ourclass$, consider the selfadjoint extensions $A_0$ and
$A_{\gamma}$, with $0<\gamma<\pi$. Then
$A$ is entire in the generalized sense if and only if $\Sp(A_0)$ and
$\Sp(A_{\gamma})$ obey conditions (C1), (C2) and (C3$\,^\flat$).
\end{theorem}

A statement analogous to Proposition~\ref{thm:krein-particular-case}
can also be formulated for generalized entire operators. We leave
the details to the reader.

\begin{remark}
  The generalized notion of entire operator discussed above may be
  extended to a notion of an operator having a generalized entire
  gauge in a suitable defined Hilbert space $\cH_{-N}$, expected to be
  the space of linear functionals of an appropriate Gelfand triplet
  associated with $\cH$. Since such space is likely to be related with
  the set of functions $N$-associated to the de Branges space
  $\widehat{\cH}$, Theorem~3.2 of \cite{woracek2} should provide a
  spectral characterization of such kind of operators. This matter
  will be treated elsewhere.
\end{remark}

\end{document}